# Running Identical Threads in C-Slow Retiming based Designs for Functional Failure Detection

T. Strauch

*Abstract*—This paper shows the usage of C-Slow Retiming (CSR) in safety critical and low power applications. CSR generates C copies of a design by reusing the given logic resources in a time sliced fashion. When all C design copies are stimulated with the same input values, then all C design copies should behave the same way and will therefore create a redundant system. The paper shows that this special method of using CSR offers great benefits when used in safety critical and low power applications. Additional optimization techniques towards reducing register count are shown and an on-the-fly recovery mechanism is discussed.

## I. INTRODUCTION

Safety critical applications use redundant designs and design state comparison techniques to detect potential design misbehavior. An example is a motor control circuit, where a malfunction of the design could generate life threatening conditions. On the other hand, a full stop and restart of a system is sometimes costly and should potentially be avoided with a very fast recovery mechanism.

Another application for using redundant designs are the control systems of satellites. Single event upsets (SEUs) must be detected before they could endanger costly missions in the orbit. It is beneficial when the power consumption of the redundant systems can also be reduced.

C-Slow Retiming (CSR) provides C copies of a given design by inserting registers and reusing the combinatorial logic in a time sliced fashion. It is already used in the 1960's, for example in the Barrel processors from the CDC 6000 series. Publications about this technique have been rare throughout the last decade. This paper shows a novel approach to use CSR-ed designs when redundant designs are needed. It concentrates on the power consumption aspect, an area reduction based on a register removal technique and it shows an on-the-fly recovery mechanism.

### A. Background

The ever increasing demands for higher performance and higher throughput of designs have led to various techniques. Lin *et al.* present in [1] an efficient retiming algorithm and in [2] a retiming algorithm for FPGAs is shown by Singh *et al.*. Retiming for wire pipelined SoC buses is discussed by Lin *et al.* in [3]. Kroening *et al.* outline automatic pipelining of designs in [4]. The pipelining of sequential circuits with wave steering is shown by Macchiarulo *et al.* in [5]. Leiserson *et al.* introduce the concept of C-Slow Retiming (CSR) in [6]. Bufistov *et al.* [7] present a formulation as a general model for retiming and recycling, which also accepts an interpretation of the CSR problem. Weaver *et al.* present the effects of CSR on 3 different benchmarks in [8] and the post-placements CSR-ing of a microprocessor on an FPGA [9]. Baumgartner *et al.* [10] present an abstraction algorithm for the verification of generalized C-slow designs. In recent publications, CSR is used to maximize the throughout-area efficiency in [11] by Su *et al.* and on simultaneous multithreading processors in [12] by Akram *et al.*.

### B. Contribution and Paper Organization

To the best of the author's knowledge, power consumption (P) has not been considered in publications about C-Slow Retiming (CSR). The same is true for the aspect of using a CSR-ed design as a C-times redundant system.

The paper demonstrates how to use CSR for SEU detection and design state on-the-fly recovery. The method is then further developed and optimized to reduce area (FPGA utilization) and the P of the application. Results of two 32-bit processors on a low-cost FPGA are given.

Section II outlines the CSR technology. In section III the P of CSR-ed designs is discussed. A method to detect single event upsets and how it can be combined with an on-the-fly recovery mechanism is shown in section IV. Section V proposes a method to reduce the number of registers which are used in the standard CSR approach. The paper finishes with results and a summary in the sections VI and VII.

## II. C-Slow Retiming

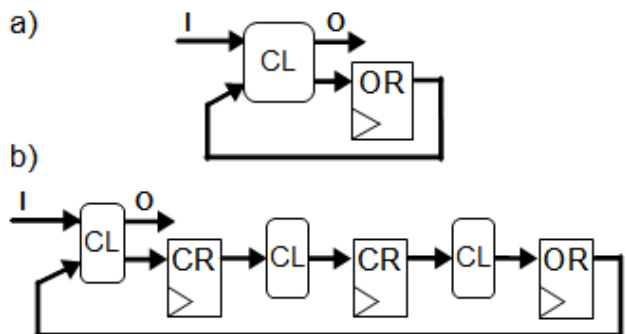

Fig. 1. a) Simplified design. b) Applying CSR technique.

### A. Theory of CSR

Fig. 1a shows the basic structure of a sequential circuit with its combinatorial logic (CL), inputs (I) and outputs (O) and original registers (OR). In Fig. 1b, the CL is sliced into three (C=3) parts, and each original path has now two (C-1) additional registers. This is the basic idea behind CSR. It results in C functionally independent design copies which use the logic in a time sliced fashion. It shows how different parts of the logic are used during different cycles. It now takes three micro-cycles to achieve the same result as in one original cycle. In Fig. 1, inputs and outputs are valid at the same time slice. The implemented register sets are called "C-Slow Retiming Registers", CRs. They are placed at different C-levels. Fig. 1b shows one basic rule of CSR. There are only paths between consecutive CRs and also from the last CRs to the original register set and from the original register set to the first CRs.

The maximum frequency of the given design (Fig. 1a) is defined as Fd and the maximum frequency of a CSR-ed design (Fig. 1b) as Fcsr, whereas:

$$Fcsr \sim Fd * C \qquad (1)$$

The individual cycle of a CSR-ed design is called a micro cycle. By generating C independent copies of the design, all running – theoretically – at Fd, it can be said that the system frequency Fsys is equal to Fcsr:

$$Fsys = Fcsr \sim Fd * C \qquad (2)$$

In theory, this could lead to an unlimited performance increase. Evidently this cannot be done endlessly and register insertion becomes inefficient for higher C again. The results section at the end of this paper shows examples for that.

When the CSR-ed design is embedded in a certain system architecture, it is sometimes possible to remove CRs by pushing them out the inputs/outputs, before connecting the CSR-ed design with the surrounding logic.

In the remainder of this paper, processors are used to demonstrate the effectiveness of CSR, but the method is not limited to processors only. Nevertheless, the word "thread" is used synonym for the execution of a processor program or the execution of an algorithm on a digital design.

### B. CSR on RTL

CSR clearly changes the behavior of the design and can only be fully utilized when the CSR-ed core is embedded in a new logic environment. This can be done by using a multiplexer structure (for inputs). Sometimes a direct connection to registers without a multiplexer structure is doable as well (outputs). Memories are usually accessed by adding the thread identifier as MSB to the address bus.

These modifications have a dramatic impact on the design flow. It is of great advantage to have a solution on a higher level such as RTL. The CSR-ed version must be used as a new subdesign in the design and verification process. A technique has been demonstrated, which automatically modifies the design to enable CSR on RTL by Strauch in [13]. The results given in this paper are based on this technology.

### C. Verification of CSR Design Modifications

It is non-trivial to verify the correctness of CSR based designs. It looks easy on paper, but unless there is no special tool for that, the task remains critical. To the best of the author's knowledge, there is no publicly available tool for this tasks. Alternatively, static timing analysis (STA) can be used for that. When each C-level gets its own clock tree, only paths from one C-level to the next one exist. Additional paths exist from the last C-level to the original registers and from the original registers to the first C-level. It can be checked during a stand alone design level synthesis and STA run, if additional paths exist, which should not exist. The STA verifies the correctness of the register insertion process. The individual clocks can then be connected to a single clock again.

## III. Power Consumption of CSR-ed Designs

### A. Overview

It has been demonstrated, that register insertion (or pipelining) can reduce the power consumption (P) of a design. For example Lim *et al.* use a power-aware flip-flop insertion with shifted-phase clocks in [14]. The assumption that CSR also reduces the P, because it is using register insertion is not necessarily true. Empirical data based on two processors reported in the result section show, that CSR actually increases the P of a design copy significantly compared to the original design.

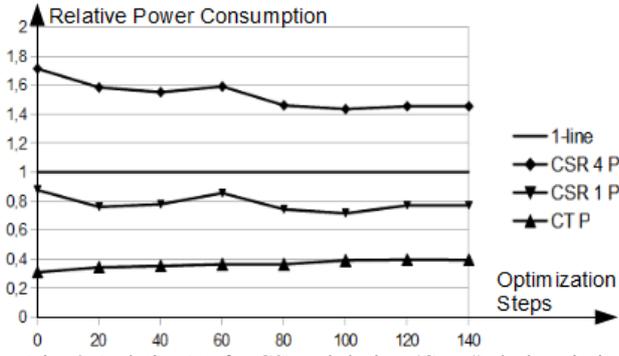

Fig. 2. Relative P of a CSR-ed design (C = 4) during timing optimization process.

When using CSR, registers have to be inserted in all existing logic paths, which also includes the feedback loops of registers. This usually adds the majority of additional registers. To achieve a reasonable timing, the CRs must be placed first of all timing-driven before power-aware register balancing technique can be applied. Nevertheless, the number of registers, which are added by CSR is high for large designs and high C's. When using CSR, the number of resulting registers is less or roughly C-times higher.

Two (out of many) sources for power consumption (P) in digital designs are the clock tree activity and the switching activity of the combinatorial logic, which also causes glitches in the design. In the conventional CSR approach, a design is instantiated N-times. Therefore, the number of resulting registers is N-times higher, but the clock speed remains the same. In short: **"N-times more registers running at the original speed"**. When using CSR, the number of registers is roughly C-times higher, but the clock speed must also be C-times higher to achieve the same performance. In short: **"C-times more registers running at C-times of the original speed"**. This results in a C-times higher P of the clock tree of the CSR-ed designs compared the one of the alternative approach to instantiate individual designs (assuming N == C). Therefore, CSR only reduces the P when the P increase due to the higher clock tree activity is less than the P savings that can be realized with glitch reduction by register insertion.

The CSR algorithm used in this paper places registers at the end of each path and then moves the individual CRs throughout the combinatorial logic until the best timing is achieved (timing optimization process). Further power aware register balancing techniques are not applied, because this paper concentrates on a register removal technique, which is demonstrated later.

In Fig. 2 the "CSR 4 P" line shows the relative P of one thread compared to the P when running the thread on the original core ("1-line"). In case of the "CSR 4 P" line, 4 different threads are executed but only the average P per thread is used. It starts with 71% P overhead at the beginning of the timing optimization process. This is due to the facts, that the signals generate toggling activity when passing through the additional registers and that the higher register load (and clock frequency) generate a higher clock tree P. The P overhead drops from 71% to 45% during the timing optimization process when a better register distribution throughout the logic – mainly on the datapath - is reached. It can be argued, that this P reduction comes from the fact that the number of longer logic paths is reduced and therefore the probability to generate power consuming signal glitches is reduced. To combine this timing driven approach with power aware optimization techniques (as shown in [14]) is outside the scope of this paper.

Fig. 2 shows the relative P of the clock tree compared to the P of the original thread during the timing optimization process ("CT P"). The relative P per thread of the clock tree increases due to the rising number of registers when improving the timing of a CSR-ed design. The line "CSR 1 P" shows the P of a single thread when only identical threads are executed. This will be discussed in the next section.

The power consumed by a CMOS integrated circuit in the quiescent state (when the circuit is not switching and inputs are held at static values) is commonly called leakage power. This current is presumably higher for larger designs. When CSR is used on an ASIC, it can be argued, that the smaller CSR-ed design consumes less leakage power compared to the larger design of the alternative approach to instantiate individual designs.

B. *Using both Clock Edges in CSR*

Another special CSR approach is to use inverted clock edges for every other C-level. This approach makes only sense when an even number of design slices exists (C = 2, 4, …). The number of resulting design copies will be half of the design slices C / 2. In this case, the P for each thread did not change significantly. A group of threads can then be used for SEU detection, as discussed in the next section.

C. *P when Running Identical Threads*

In Fig. 2 it can be seen how the P changes when applying the CSR algorithm (C = 4) on a given example design and identical threads are executed ("CSR 1 P" line). In this case, the average P of a single thread is in the range of 87% to 77% of the P generated by the same thread executed on the original design. Here the P of the clock tree increases due to the higher register count and the higher clock frequency, but only one thread generates switching activity, so that the average P of a single thread is less than the one of the same thread on the original design. Why it can be useful to execute identical threads will be explained in the next section.

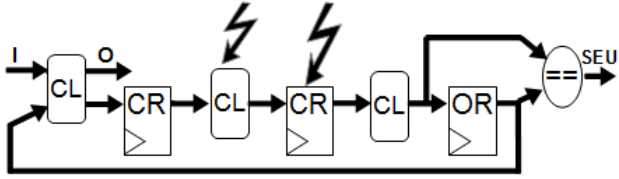

Fig. 3. Comparing signal values at key registers to detect an SEU.

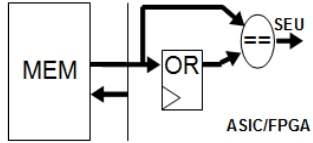

Fig. 4. Detecting faulty memory content with (interleaved) memory access to identical data at different memory locations.

## IV. DETECTING A SINGLE EVENT UPSET (SEU) USING CSR

### A. Detecting an SEU with Standard CSR

One way to detect a single event upset (SEU) is the duplication of a design (redundancy) and to compare key registers and/or outputs. When an SEU occurs, at least one design runs different and further actions can be taken. CSR supports this feature when executing (a group of) identical threads. In Fig. 3, all threads execute the same algorithm (or program) and use the logic in a time shared fashion. Therefore only a limited number of threads are affected when an SEU occurs. Multiple identical threads are most likely affected differently because each one of them is in a different design state. When this difference affects the state of key registers, it can be detected by a support logic which compares the states of consecutive threads. A mismatch indicates that threads run differently.

This method was tested on two different processors using error injection techniques in simulation (as discussed by Braza et al. in [15]) to verify the behavior. The comparison logic can also be pipelined so that the timing of the original circuit is not impacted. Input multiplexing becomes obsolete when identical threads are executed.

### B. Handling Memories

The proposed method is not limited to processors only. All kinds of digital designs and subsystems (including the system bus, peripherals, accelerators, etc.) can be elements of the CSR modifications. Although the simplification in Fig. 1 is still valid, it is obvious that memories are usually duplicated in CSR-ed designs. A processor's register file will have a C-times larger memory space than the one of the original design. The thread identifier is then added as most significant bits (MSB) to the memory address bus. When applicable, larger external data or program memories can also be duplicated, and each C-copy of the design gets exclusive access to its specific sections. In case of running identical threads, the thread identifier can also be added as least significant bits (LSB) to the address bus in order to support potential memory burst features.

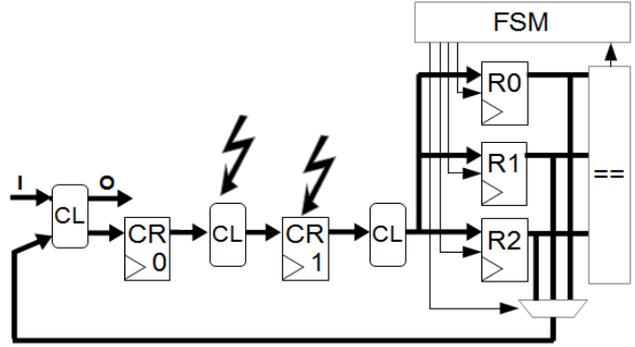

Fig. 5. On-the-fly recovery.

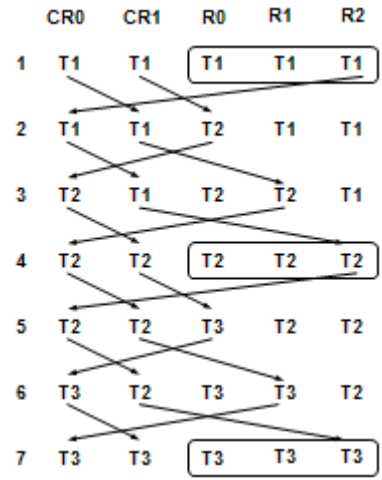

Fig. 6. Design copy propagation.

With the memory duplication, it is guaranteed, that a faulty memory content will only impact a single design copy, whereas the remaining design copies are not affected. Fig. 4. shows, that when an instruction or data from an external memory reaches the incoming register with a comparison logic, faulty data or instructions are detected by the proposed mechanism. Once a thread behaves differently, it can be recovered by using the method discussed in the next section, which can also help to clear spikes from incoming datastreams.

### C. Recovery

When an SEU is detected, safety critical designs can restart or execute predefined software recovery routines. When using CSR, an on-the-fly recovery is possible. Fig. 5 shows the CSR-ed design enhanced by an SEU detection circuit. When C >= 3, the SEU detection circuit uses a majority decoder to detect the failing thread by comparing the key register values of C identical threads. This is done every C micro-cycles.

It will be shown in detail, how a modified write enable sequence - controlled by a finite state machine (FSM) - then overwrites the specific Rn register associated with the failing thread. This write control must also be combined with a specific Rn read sequence to establish an on-the-fly recovery mechanism.

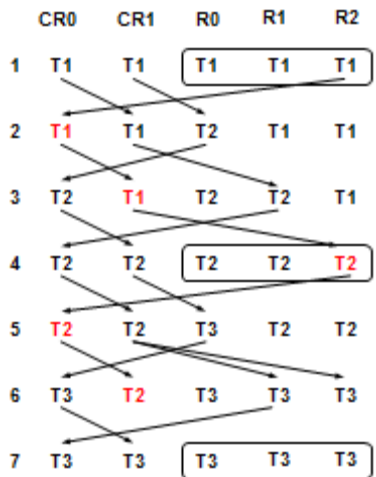

Fig. 7. Fault detection at R2.

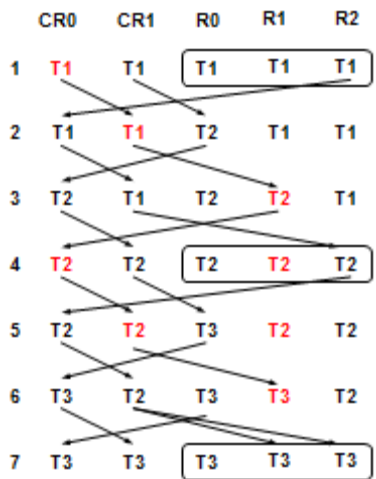

Fig. 8. Fault detection at R1.

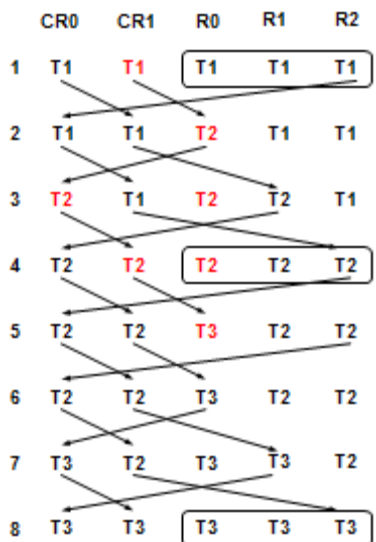

Fig. 9. Fault detection at R0.

In Fig. 6 the propagation of the individual design copies is shown. In cycle 2, CR0 takes over the state of R2, CR1 the state of CR0 and R0 is update with a new state values. In cycle 3, CR0 takes over the state of R0 and in cycle 4 the state of R1. R1 takes over the state of CR1 in cycle 3 and R2 the one of CR1 in cycle 4. This mechanism is repeated continuously. Every third (C-ed) cycle (1, 4, 7, ...) the Rn values are compared.

Assuming that in Fig 7, R2 is detected as faulty in cycle 4 (marked red). This means, that in cycle 2 and cycle 3 the register could have already been faulty and that this faulty value might have already been propagated through the CR0 and CR1 registers. In order to overwrite that design copy with a valid state, the T2 design state in cycle 6 is not only copied to R1, but also to R2, which will then not be overwritten in cycle 7.

Overwriting a failing design state works almost the same way when the design copy at R1 is detected as faulty in cycle 4. In that case (Fig. 8), the incoming value for T3 in cycle 7 is not only copied to R2, but also to R1. CR0 takes over the value of R0 in cycle 7, and not the one of R1.

Clearing a faulty R0 state is more tricky (Fig. 9). Assuming that this case is detected in cycle 4. To solve that situation, a delay cycle (5) is inserted and the design state T2 is merged into the CR-line again (CR0 takes over the values from C2 in two consecutive cycles 5 and 6). When the next comparison occurs (cycle 8) the copies will have identical states again.

When signals leave the system or travel from a CSR-ed design section to a non-CSR-ed design, faulty signal behavior resulting from a faulty design copy can be eliminated by using a majority decoder. As already mentioned, the recovery mechanism can also be used for incoming datastreams.

It is important to notice, that this recovery mechanism only uses an additional hold signal for each Rn and one additional "read" multiplexer. The Rn write mechanism can easily be combined with a gated clock structure of the clock tree when implemented on an ASIC.

The logic for the recovery mechanism is pipelined, because CSR-ed designs tend to be timing critical. It uses one cycle to do the comparison, and the removal of a faulty state is executed in one of the following cycles.

The technique has been successfully simulated on RTL using a simple 1-out-of-3 majority decoder and an error injection mechanism. The results were always a full design on-the-fly recovery. The area overhead of this approach is reported in the result section. This very fast on-the-fly recovery mechanism is almost impossible to achieve when using the standard SEU detection concept of individual redundant design implementations.

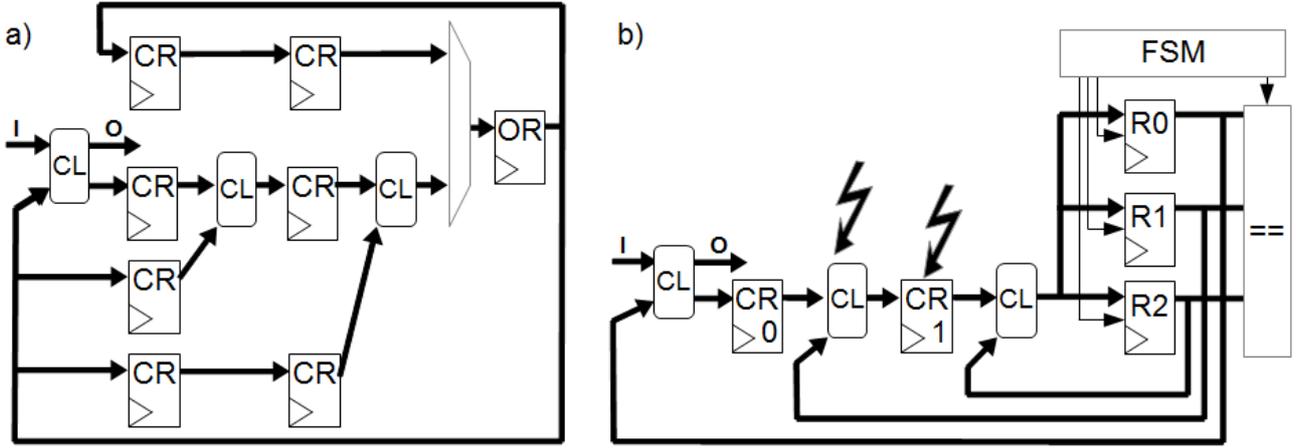

Fig. 10. a) Shift Registers generated by Register Feedback Loops and adjacent C-Slow Retiming Registers (CRs). b) CSR-ed design with SEU detection circuit and a minimal set of C-Slow-Retiming Registers (CRs).

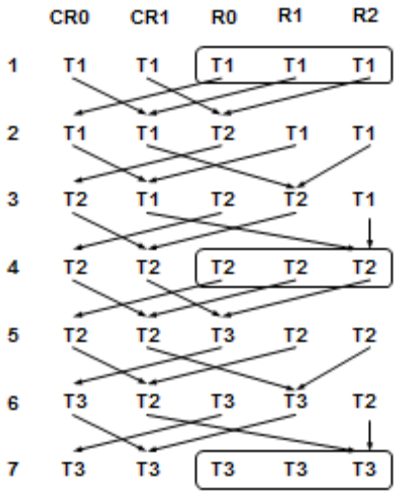

Fig. 11. Design copy propagation.

## V. CSR-ed Designs with minimal Register Count

Fig. 10a shows a design after applying the CSR method. It is essentially the same schematic as already shown in Fig. 1b, but this time CRs which are directly connected (and build shift registers) are singled out. It can be seen that CSR generates a high number of shift registers by adding registers to the feedback loop of the original registers. Additional shift registers are generated on the paths through the combinatorial logic. The CRs contribute to the majority of area and P increase.

When identical threads are executed, the number of shift registers can be reduced by using a modified CSR algorithm, called "CSR minimal" (CSRmin). Fig. 10b shows, how the CSR-ed design can be improved to reduce the number of CRs. When using CSRmin, the outputs of the Rns drive the combinatorial logic at different C-levels, so that shift registers generated by consecutive CRs can be removed by connecting the combinatorial logic with the relevant Rn. The Rn write mechanism is controlled by a FSM.

Fig. 11. shows the design copy propagation for multiple cycles. CR0 is always updated by using R0 and CR1 is using CR0 and R1. The Rn register are continuously updated (R0, R1, R2, R0, …) by using CR1 and R2 as the source. Again, every third (C'd) cycle the register values can be compared to detect potential design thread mismatch.

This method does have a positive impact on the overall register count (area) and the P of the CSR-ed design running identical threads improves. Empirical data on 2 different processors is shown in the result section.

It is also possible to eliminate only the shift registers generated by the feedback loop and to update the original register (OR) every C'd cycle. In this case, the OR does not have to be duplicated (generating Rns) and the read mechanism works like in Fig. 3. This approach is not investigated any further, because this paper concentrates on SEU detection and possible recovery mechanism.

The CR removal method discussed in this section is fundamentally different to the CR removal task which is mentioned in "II. C-Slow Retiming, A. Theory of CSR". When running identical threads, **internal** CRs can be removed as well, whereas when different threads are used, only CRs at the input/outputs of the CSR-ed design can potentially be eliminated.

Not all faulty changes of all registers at any given time can be detected, when the CSRmin method is enhanced by a recovery mechanism. When R2 flips its value in cycle 2, the effort to catch the fault and to recover from it is impossible. Although most defects can be removed, the recovery mechanism is not added to CSRmin in this paper.

TABLE 1. RESULTS FOR ARM3 CORE, PART I

| C | Registers | | Occupied Slices | | Performance [ps] | | | PpA | P [mW] | P same [mW] | | | P diff [mW] | | |
|---|---|---|---|---|---|---|---|---|---|---|---|---|---|---|---|
| | | rel to 1 | | rel to 1 | | rel to 1 | rel to A | | | | @ perf. | @ max | | @ perf. | @ max |
| 1 | 670 | | 825 | | 18250 | | | 66,42 | 22,1 | 22,1 | | | n.a. | | |
| 2 | 1683 | 251% | 1015 | 123% | 11850 | 154% | 77% | 125% | 44,2 | 22,79 | 52% | 67% | 50,05 | 113% | 147% |
| 3 | 1768 | 264% | 1018 | 123% | 8917 | 205% | 68% | 166% | 66,3 | 35,00 | 53% | 77% | 66,72 | 101% | 148% |
| 4 | 2091 | 312% | 1029 | 125% | 7210 | 253% | 63% | 203% | 88,4 | 41,01 | 46% | 73% | 81,50 | 92% | 146% |
| 5 | 2211 | 330% | 1177 | 143% | 6234 | 293% | 59% | 205% | 110,5 | 43,91 | 40% | 68% | 94,27 | 85% | 146% |

TABLE 2. RESULTS FOR THE OR1200 CORE, PART I

| C | Registers | | Occupied Slices | | Performance [ps] | | | PpA | P [mW] | P same [mW] | | | P diff [mW] | | |
|---|---|---|---|---|---|---|---|---|---|---|---|---|---|---|---|
| | | rel to 1 | | rel to 1 | | rel to 1 | rel to A | | | | @ perf. | @ max | | @ perf. | @ max |
| 1 | 1280 | | 994 | | 14008 | | | 71,82 | 42,4 | 42,4 | | | n.a. | | |
| 2 | 2741 | 214% | 1254 | 126% | 9080 | 154% | 77% | 122% | 84,8 | 43,55 | 51% | 67% | 126,50 | 149% | 193% |
| 3 | 3573 | 279% | 1335 | 134% | 7127 | 197% | 66% | 146% | 127,2 | 54,58 | 43% | 65% | 163,14 | 128% | 196% |
| 4 | 3901 | 305% | 1316 | 132% | 6334 | 221% | 55% | 167% | 169,6 | 64,18 | 38% | 68% | 174,71 | 103% | 186% |
| 5 | 4210 | 329% | 1361 | 137% | 5973 | 235% | 47% | 171% | 212 | 69,61 | 33% | 70% | 197,00 | 93% | 198% |

## VI. RESULTS

The numbers in this results section are based on two CPUs. The RTL code for the ARM3 core ("Amber", 32-bit RISC processor, 3-stage pipeline, ARM v2a) and the OR1200 ("OR1000", 32-bit scalar RISC, 5-stage pipeline, MMU, DSP capabilities, TLB, instruction and data cache) can be found at [16]. The designs are implemented on a Xilinx Spartan-6 LX16 (-2ftg256). The clock is generated externally.

A tool called "CoreMultiplier" is used in this paper. Its algorithm is described by Strauch in [13]. The original RTL codes of the 2 processor designs are taken and the tool automatically inserts CRs on RTL. This is done timing-driven. For this work, CoreMultiplier is enhanced to insert the recovery logic and to remove shift-register structures where possible.

### A. Results using Standard CSR

Table 1 shows the results of a CSR-ed ARM3 core. When multiplying the functionality by C = 2...5, the number of registers increases up to 330%. At the same time, the number of occupied slices remains relatively stable. This indicates, that the additional registers nicely fit into the already used slices. In other words, you have 5 times the functionality with just an area overhead of 43% when using CSR.

The performance increases with each C step. Although it does not reach the performance (200%, 300%, …, 500%) of the alternative concept by implementing individual processors (called "A" in the remainder of this section), it has a reasonably timing of 6.234 ns. This is a performance increase of up to 293% compared to a single core implementation ("rel to 1"), but it only reaches 59% ("rel to A") of the performance of A. Better results can be achieved with more advanced technologies like the Virtex family, as can be seen in [13], and most likely in ASIC technologies.

When a single core with 825 occupied slices can run at 18.250 ns, the performance per area (PpA) factor can be calculated to 66,42 kHz/slice (Table 1). It can be seen in the PpA column, that this factor increases by up to 205% for C = 5. In other words, when CSR can be used, more performance can be realized on a given size. Nevertheless, increasing C becomes less efficient for higher C.

The P of the original ARM3 core is 22,1 mW, running at maximum speed (18.250 ns). When instantiating individual ARM3 processors, the P multiplies accordingly. It must be distinguished between running the same program on all available designs or running different programs.

When running the **same** program at the maximum possible speed, the P decreases to 40% compared to A. This is certainly due to the fact, that the maximum possible speed is less than the one of A.

Even when the CSR-ed core could be run at the theoretical possible speed (cycle time = 18.250 ns / C), the P would only be in the range of 68% to 77% of the A. The P seams to be relatively constant and independent of C when running the same program. As can be seen in Fig. 2, the P is relatively independent of the CSR timing optimization process when moving registers throughout the combinatorial logic.

The P changes relatively to the P of the A from 113% to 85% when C is increased while running **different** programs. When running the design at the theoretical possible speed (18.250ns / C), the P is around 147% of the P of A. It turned out that this number is relatively constant for different Cs. A CSR-ed design uses less registers than A, but can run (theoretically) C times faster, which results in a higher P of the clock tree than the one of A.

TABLE 3. RESULTS FOR THE ARM3 CORE, PART II

| C | Register CSRrec | | OS CSRrec | | Register CSRmin | | OS CSRmin | | Performance CSRmin [ps] | | | P CSRrec [mW] | | P CSRmin [mW] | |
|---|---|---|---|---|---|---|---|---|---|---|---|---|---|---|---|
| 1 | 670 | rel to 1 | 825 | rel to 1 | 670 | rel to 1 | 825 | rel to 1 | 18250 | rel to 1 | rel to A | @ max | rel to 1 | @ max | rel to 1 |
| 2 | | | | | 1093 | 163% | 967 | 117% | 9428 | 194% | 97% | | | 23,36 | 53% |
| 3 | 2720 | 406% | 1444 | 175% | 1230 | 184% | 824 | 100% | 7425 | 246% | 82% | 59,46 | 90% | 41,77 | 63% |
| 4 | 2966 | 443% | 1745 | 212% | 1498 | 224% | 969 | 117% | 7337 | 249% | 62% | 74,83 | 85% | 52,46 | 59% |
| 5 | 3071 | 458% | 1989 | 241% | 1655 | 247% | 1054 | 128% | 6519 | 280% | 56% | 87,28 | 79% | 60,24 | 55% |

TABLE 4. RESULTS FOR THE OR1200 CORE, PART II

| C | Register CSRrec | | OS CSRrec | | Register CSRmin | | OS CSRmin | | Performance CSRmin [ps] | | | P CSRrec [mW] | | P CSRmin [mW] | |
|---|---|---|---|---|---|---|---|---|---|---|---|---|---|---|---|
| 1 | 1280 | rel to 1 | 994 | rel to 1 | 1280 | rel to 1 | 994 | rel to 1 | 14008 | rel to 1 | rel to A | @ max | rel to 1 | @ max | rel to 1 |
| 2 | | | | | 2035 | 159% | 1282 | 129% | 8077 | 173% | 87% | | | 58,51 | 69% |
| 3 | 6291 | 491% | 1968 | 198% | 2214 | 173% | 1332 | 134% | 6334 | 221% | 74% | 108,12 | 85% | 86,50 | 68% |
| 4 | 6349 | 496% | 2097 | 211% | 2522 | 197% | 1451 | 146% | 6334 | 221% | 55% | 137,38 | 81% | 111,94 | 66% |
| 5 | 6454 | 504% | 2734 | 275% | 2816 | 220% | 1501 | 151% | 5973 | 235% | 47% | 167,48 | 79% | 142,04 | 67% |

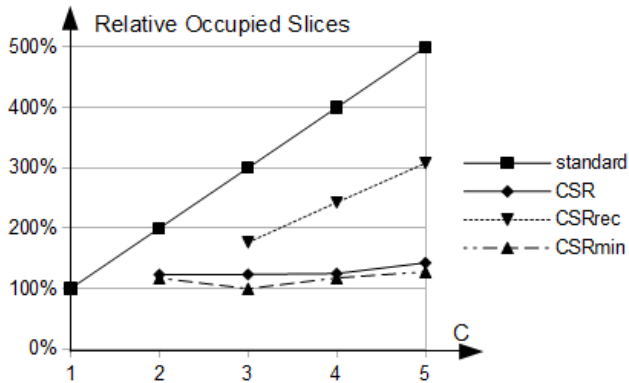

Fig. 12. Relative Occupied Slices

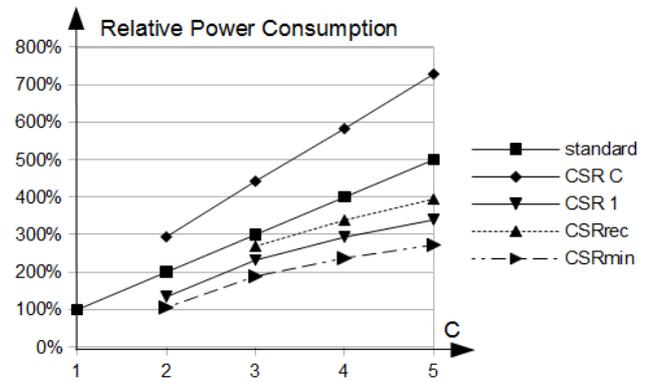

Fig. 13. Relative Power Consumption

Similar numbers can be found for the CSR-ed implementation of the OR1200 core. The relative number of registers increases to up to 329% (Table 2), whereas the number of occupied slices only reaches 137% for C=5. The performance increase is less optimal over an increasing number of copies. This is due to the already fast cycle time of the original core and the relatively slow technology (Spartan 6). Better results can be achieved on a more advanced technology (Virtex 5), as reported in [13]. The P of the original core is 42,4 mW (Tables 6). The P when running the same or different programs and with increasing numbers of copies is listed as well. When running the same thread and removing obsolete shift registers, the area increase is only 11%.

### B. Results using CSRmin, SEU Detection and Recovery

The results for C-Slow Retimed designs using SEU detection based on the standard CSR or the CSRmin algorithm are shown in the Table 3 (ARM3) and Table 4 (OR1200). Fig. 12 shows the results of the ARM3 graphically. It can be seen, that the additional registers needed for a CSR based recovery (Table 3, column 2) occupy for C = 3 only 76% more area (occupied slices (OS)) and 208% more OS for C = 5.

The best results in terms of OS can be achieved by using the CSRmin algorithm. In this case, a triple CSR-ed ARM3 processor can be implemented with no area overhead (Table 3, column 9). When C = 5, the CSRmin algorithm only generates 28% overhead compared to the original design. The alternative concept of implementing 5 individual processors would generate an area overhead of 400% of the original design.

The results show, that CSR designs can be perfectly packed into FPGAs. The CSRmin algorithm further improves the number by removing a majority of CRs. By doing that, CSRmin can achieve a reasonable better performance closer to A for C = 2 and C = 3 compared to CSR. This can be seen in Table 3 and Table 4, "Performance CSRmin [ps]" column compared to the same column in the CSR related Table 1 and 2.

In the Tables 1 and 3 as well as in Fig. 13 the P of the individual CSR solutions is shown. For the 2 empirical data sets, the P increases dramatically when different threads are executed ("CSR C"). When running the same thread, the P per thread is in favor of CSR compared to A. Best P related results can be achieved with the CSRmin algorithm. Adding the recovery mechanism increases the P for the CSR solutions again, but it still below the P of A for the two testcases.

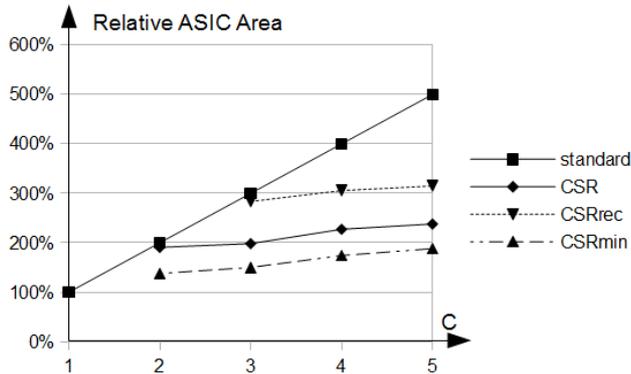
Fig. 14. Relative ASIC Area

### C. Projected Results for ASICs

This section projects the results of the FPGA based synthesis flow on ASICs. This is done by setting the area ratio of registers and combinatorial logic to 60/40. This ratio is based on a synthesis run using an ASIC library and the OR1200. The area of the combinatorial logic is estimated to be roughly the same and the register count can be seen in the Tables shown so far.

In Fig. 14 it can be seen, that the CSR, CSRrec and CSRmin methods decrease the relative ASIC area compared to A. In case of the CSRmin algorithm, an SEU can be detected (C = 2) with just 37**% ASIC area overhead of the original design** and singled out (C = 3) with just 50% area overhead.

The CSRrec algorithms adds a considerable amount of additional registers, which increases the area consumption of ASICs dramatically, compared to the FPGA solution. Nevertheless, the CSR-ed design with recovery mechanism does not generate considerable more area than A and has even less area on the given ASIC projections.

In terms of P, it can be assumed, that the results are analog to those of the FPGA results. The registers added to support the recovery mechanism can use gated clock trees, which further reduces the P on ASICs. It is also doable to create a specific library cell with C registers to further reduce power and area consumption on ASICs. This library cell can then be enhanced with a comparison logic and/or an output multiplexer.

## VII. Summary

### A. Performance

In general it can be said, that an individual thread runs most likely slower on an CSR-ed design compared to its execution on the original design. This disadvantage can be reduced by using more advanced technologies [13]. Nevertheless, CSR improves the performance per area factor.

### B. C-Slow Retiming

On FPGAs, the multithreaded CSR solution needs less occupied slices, due to the high number of available registers on FPGAs. It has been shown, that CSR based register insertion reduces the P (Fig. 2), but the increased clock tree activity of CSR-ed designs has a grater negative impact on the P. When running different threads, each thread consumes roughly 40% more power than the alternative approach to instantiate individual designs. Whereas when identical threads are executed, the power consumption is in favor of CSR, because a thread consumes 30% less power on an CSR-ed design than on the original core implementation.

### C. C-Slow Retiming Recovery

For CSR an on-the-fly recovery mechanism is shown. It increases the area and power consumption compared to the standard CSR solution, but it is still advantageous to the alternative approach to instantiate individual cores, where a single cycle recovery is impossible to implement.

### D. C-Slow Retiming Reduced

The area and power consumption can be further improved when the CSRmin technique is used, which eliminates a majority of the registers inserted by CSR. For FPGAs, empirical data show that an SEU detection with C = 2 comes at almost no additional area costs or additional power consumption and only a minor performance penalty compared to the original design, whereas the alternative concepts needs twice the area and consumes twice the power.

A combined solution of the standard design duplication method and the CSRmin algorithm can be beneficial. In this case both designs are enhanced with the CSRmin method (C = 2) at almost no additional area cost and no additional P, but it would be easy to identify, which one of the two designs is faulty.

### E. CSR, CSRrec and CSRmin on ASICs

It looks promising to use this method on ASICs and design implementations, where SEU detection, power consumption and design area play an important role. The CSRmin method reduces the area by 25% (C = 2), 50% (C = 3) or even more for higher C's. The P can be assumed to be in favor of the proposed methods, as shown for FPGAs. The reduced area, a reduced Iddq and ASIC specific features like gated clock trees further reduce the P. With the proposed CSRrec method a design state recovery mechanism (C = 3) becomes available with no additional area costs compared to the alternative implementation.

### F. Comparison to Alternative Concepts

To the best of the author's knowledge, no literature exists, which provides reasonable data for comparison. To compare the proposed methods with alternative concepts, area, power consumption and the behavioral modifications (multiple design copies) must be considered at the same time. The results are all compared to the most obvious alternative to instantiate individual designs. Most concepts which only target one aspect (like power reduction in [14]) can still be applied on top of the method discussed in this paper.

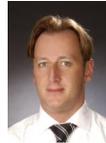

**Tobias Strauch** received his Diploma (FH) at the University of applied science (FH) Furtwangen, Germany in '98. He works for EDAptix in Munich, Germany. His field of interests are hardware assisted verification, TLM, C-Slow Retiming, System Hyper Pipelining, High Level ATPG, FPGA debugging and wave based data transfer.